


\font\titlefont = cmr10 scaled\magstep 4
\font\sectionfont = cmr10
\font\littlefont = cmr5
\font\eightrm = cmr8

\def\ss{\scriptstyle}
\def\sss{\scriptscriptstyle}

\newcount\tcflag
\tcflag = 0  

\ifnum\tcflag = 0 \magnification = 1200 \fi  

\global\baselineskip = 1.2\baselineskip
\global\parskip = 4pt plus 0.3pt
\global\abovedisplayskip = 18pt plus3pt minus9pt
\global\belowdisplayskip = 18pt plus3pt minus9pt
\global\abovedisplayshortskip = 6pt plus3pt
\global\belowdisplayshortskip = 6pt plus3pt

\def\barsoff{\overfullrule=0pt}


\def\endignore{}
\def\ignore #1\endignore{}

\newcount\dflag
\dflag = 0


\def\monthname{\ifcase\month
\or January \or February \or March \or April \or May \or June%
\or July \or August \or September \or October \or November %
\or December
\fi}

\newcount\dummy
\newcount\minute  
\newcount\hour
\newcount\localtime
\newcount\localday
\localtime = \time
\localday = \day

\def\advanceclock#1#2{ 
\dummy = #1
\multiply\dummy by 60
\advance\dummy by #2
\advance\localtime by \dummy
\ifnum\localtime > 1440 
\advance\localtime by -1440
\advance\localday by 1
\fi}

\def\settime{{\dummy = \localtime%
\divide\dummy by 60%
\hour = \dummy
\minute = \localtime%
\multiply\dummy by 60%
\advance\minute by -\dummy
\ifnum\minute < 10
\xdef\spacer{0} 
\else \xdef\spacer{}
\fi %
\ifnum\hour < 12
\xdef\ampm{a.m.} 
\else
\xdef\ampm{p.m.} 
\advance\hour by -12 %
\fi %
\ifnum\hour = 0 \hour = 12 \fi
\xdef\timestring{\number\hour : \spacer \number\minute%
\thinspace \ampm}}}



\def\endtitle{}
\def\title#1\endtitle{\vskip.5in\titlefont
\global\baselineskip = 2\baselineskip
#1\vskip.4in
\baselineskip = 0.5\baselineskip\rm}

\def\endauthors{}
\def\authors#1\endauthors{#1}

\def\endabstract{}
\def\abstract#1\endabstract{\vskip .3in%
\centerline{\sectionfont\bf Abstract}%
\vskip .1in
\noindent#1}

\def\nopageonenumber{\footline={\ifnum\pageno<2\hfil\else
\hss\tenrm\folio\hss\fi}}  

\newcount\nsection
\newcount\nsubsection

\def\section#1{\global\advance\nsection by 1
\nsubsection=0
\bigskip\noindent\centerline{\sectionfont \bf \number\nsection.\ #1}
\bigskip\rm\nobreak}

\def\subsection#1{\global\advance\nsubsection by 1
\bigskip\noindent\sectionfont \sl \number\nsection.\number\nsubsection)\
#1\bigskip\rm\nobreak}

\def\topic#1{{\medskip\noindent $\bullet$ \it #1:}}
\def\endtopic{\medskip}

\def\appendix#1#2{\bigskip\noindent%
\centerline{\sectionfont \bf Appendix #1.\ #2}
\bigskip\rm\nobreak}


\newcount\nref
\global\nref = 1

\def\therefs{}


\def\ref#1#2{\xdef #1{[\number\nref]}
\ifnum\nref = 1\global\xdef\therefs{\item{[\number\nref]} #2\ }
\else
\global\xdef\oldrefs{\therefs}
\global\xdef\therefs{\oldrefs\vskip.1in\item{[\number\nref]} #2\ }%
\fi%
\global\advance\nref by 1
}

\def\listrefs{\vfill\eject\section{References}\therefs}


\newcount\nfoot
\global\nfoot = 1

\def\foot#1#2{\xdef #1{(\number\nfoot)}
\footnote{${}^{\number\nfoot}$}{\eightrm #2}
\global\advance\nfoot by 1
}


\newcount\nfig
\global\nfig = 1
\def\thefigs{} 

\def\figure#1#2{\xdef #1{(\number\nfig)}
\ifnum\nfig = 1\global\xdef\thefigs{\item{(\number\nfig)} #2\ }
\else
\global\xdef\oldfigs{\thefigs}
\global\xdef\thefigs{\oldfigs\vskip.1in\item{(\number\nfig)} #2\ }%
\fi%
\global\advance\nfig by 1 } 

\def\fig#1{\xdef #1{(\number\nfig)}
\global\advance\nfig by 1 } 


\newcount\cflag
\newcount\nequation
\global\nequation = 1
\def\eqlabel{(1)}

\def\nexteqno{\ifnum\cflag = 0
\global\advance\nequation by 1
\fi
\global\cflag = 0
\xdef\eqlabel{(\number\nequation)}}

\def\lasteqno{\global\advance\nequation by -1
\xdef\eqlabel{(\number\nequation)}}

\def\label#1{\xdef #1{(\number\nequation)}
\ifnum\dflag = 1
{\escapechar = -1
\xdef\draftname{\littlefont\string#1}}
\fi}

\def\clabel#1#2{\xdef\eqlabel{(\number\nequation #2)}
\global\cflag = 1
\xdef #1{\eqlabel}
\ifnum\dflag = 1
{\escapechar = -1
\xdef\draftname{\string#1}}
\fi}

\def\cclabel#1#2{\xdef\eqlabel{#2)}
\global\cflag = 1
\xdef #1{\eqlabel}
\ifnum\dflag = 1
{\escapechar = -1
\xdef\draftname{\string#1}}
\fi}


\def\eeq{}

\def\eqnn #1\eeq{$$ #1 $$}

\def\eq #1\eeq{
\ifnum\dflag = 0
{\xdef\draftname{\ }}
\fi 
$$ #1
\eqno{\eqlabel \rlap{\ \draftname}} $$
\nexteqno}







\def\eqa #1\eeq{
\ifnum\dflag = 0
{\xdef\draftname{\ }}
\fi 
$$ \eqalignno{ #1 } $$
\global\cflag = 0}


\def\ie{{\it i.e.\/}}

\def\cf{{\it c.f.\/}}


\def\cmp#1#2#3{{\it Comm.\ Math.\ Phys.} {\bf #1} (19#2) #3}

\def\nci#1#2#3{{\it Nuovo Cimento} {\bf #1} (19#2) #3}
\def\npb#1#2#3{{\it Nucl.\ Phys.} {\bf B#1} (19#2) #3}
\def\plb#1#2#3{{\it Phys.\ Lett.} {\bf #1B} (19#2) #3}

\def\prb#1#2#3{{\it Phys.\ Rev.} {\bf B#1} (19#2) #3}

\def\prd#1#2#3{{\it Phys.\ Rev.} {\bf D#1} (19#2) #3}
\def\pr#1#2#3{{\it Phys.\ Rev.} {\bf #1} (19#2) #3}

\def\prl#1#2#3{{\it Phys.\ Rev.\ Lett.} {\bf #1} (19#2) #3}


\global\nulldelimiterspace = 0pt



\def\frac#1#2{{{#1} \over {#2}}\,}  
\def\hf{{1\over 2}}



\def\Square{{\vbox {\hrule height 0.6pt\hbox{\vrule width 0.6pt\hskip 3pt
        \vbox{\vskip 6pt}\hskip 3pt \vrule width 0.6pt}\hrule height 0.6pt}}}
\def\Dsl{\hbox{/\kern-.6700em\it D}} 
\def\dsl{\hbox{/\kern-.5300em$\partial$}}
\def\pxpsl{\hbox{/\kern-.5600em$p$}}
\def\ssl{\hbox{/\kern-.5300em$s$}}
\def\epssl{\hbox{/\kern-.5100em$\epsilon$}}
\def\delsl{\hbox{/\kern-.6300em$\nabla$}}
\def\lxpsl{\hbox{/\kern-.4300em$l$}}
\def\elxpsl{\hbox{/\kern-.4500em$\ell$}}
\def\kxpsl{\hbox{/\kern-.5100em$k$}}
\def\qxpsl{\hbox{/\kern-.5000em$q$}}
\def\sla#1{\raise.15ex\hbox{$/$}\kern-.57em #1}



\def\roughly#1{\mathrel{\raise.3ex
\hbox{$#1$\kern-.75em\lower1ex\hbox{$\sim$}}}}





\def\Scd{{\cal D}}

\def\Scl{{\cal L}}


\def\ssb{{\sss B}}

\def\ssf{{\sss F}}
\def\ssg{{\sss G}}

\def\ssl{{\sss L}}







\nopageonenumber
\baselineskip = 18pt
\barsoff


\def\psibar{\overline{\psi}}

\def\MSbar{\overline{MS}}


\line{hep-th/9407078 \hfil McGill-94/33, NEIP-94-006, OSLO-TP 10-94}
\vskip .3in

\title
\centerline{Bosonization in Higher Dimensions}
\endtitle

\vskip 0.2in
\authors
\centerline{C.P. Burgess,${}^{a,b*}$ C.A. L\"utken${}^c$ and F.
Quevedo${}^{a*}$}
\vskip .2in
\centerline{\it ${}^a$ Institut de Physique, Universit\'e de Neuch\^atel}
\centerline{\it CH-2000 Neuch\^atel, Switzerland.}
\vskip .1in
\centerline{\it ${}^b$ Physics Department, McGill University}
\centerline{\it 3600 University St., Montr\'eal, Qu\'ebec, Canada, H3A 2T8.}
\vskip .1in
\centerline{\it ${}^c$ Physics Department, University of Oslo}
\centerline{\it P.O. Box 1048, Blindern, N-0316, Oslo, Norway.}
\endauthors

\footnote{}{\eightrm * Research supported by the Swiss National Foundation.}


\abstract
\vbox{\baselineskip 15pt
Using the recently discovered connection between bosonization and
duality transformations, we give an explicit path-integral representation
for the bosonization of a massive fermion coupled to a $U(1)$ gauge potential
(such as electromagnetism) in $d \ge 2$ space ($D=d+1 \ge 3$
spacetime) dimensions. We perform this integral explicitly in the limit of
large fermion mass. We find that the bosonic theory is described by a
rank $d-1$ antisymmetric Kalb-Ramond-type gauge potential, whose
action is local for $d=2$ (given by a Chern-Simons action),
but nonlocal for $d \ge 3$. By coupling to
a statistical Chern-Simons field for $d=2$, we obtain a bosonized
formulation of anyons. The bosonic theory may be further dualized to a
theory involving purely scalars, for any $d$, and we show this
to be a higher-derivative theory for which the scalar decouples from
the $U(1)$ gauge potential.  }
\endabstract


\vfill\eject
\section{Introduction}

\ref\linearbos{A. Luther and I. Peschel, \prb{9}{74}{2911};
S. Coleman, \prd{11}{75}{2088};
S. Mandelstam, \prd{11}{75}{3026}}
\ref\higherdim{A. Luther, \prd{19}{79}{320};
H. Aratyn, \npb{227}{83}{172};
A. Luther and K.D. Schotte, \npb{242}{84}{407};
F.~D.~M.~Haldane, Helv.~Phys.~Acta. {\bf 65} (1992) 152;
A.~Houghton and B.~Marston, \prb{48}{93}{7790};
A.~H.~Castro Neto and E.~H.~Fradkin, \prl{72}{94}{1393};
\prb{49}{94}{} (to appear).}
The technique of bosonization consists of the replacement of a known
system of fermions with a theory of bosons which has a {\it completely
equivalent} physical content, including identical spectra and
interactions \linearbos. It provides an extremely useful tool for
analyzing such fermionic systems, since it permits the application
to them of powerful techniques that have been developed for bosonic
systems. A major limitation of the bosonization technique, however, is its
present utility only in $d = 1$ space dimension (\ie\ $D = d+1 = 2$
spacetime dimensions). This is in spite of a number of efforts
\higherdim\ to extend the theory of bosonization to higher dimensions.

\ref\abelbosedual{C.P. Burgess and F. Quevedo, \npb{421}{94}{373}.}
\ref\damgaardrefs{For an approach which is similar to ours in spirit see:
P.H. Damgaard, H.B. Nielsen and R. Sollacher,
\npb{385}{92}{227}; \plb{296}{92}{132}; \npb{414}{94}{541};
\plb{322}{94}{131}; preprint CERN-TH-7347/94 (hep-th/9407022).}
\ref\nonlinearbos{E. Witten, \cmp{92}{84}{455}.}
\ref\nonabelbosedual{C.P. Burgess and F. Quevedo, \plb{329}{94}{457}.}
\ref\doq{X. de la Ossa and F. Quevedo, \npb{403}{93}{377}.}
The purpose of this note is to present a different approach to
bosonization in dimensions $d \ge 2$. Our approach is based upon the
recently-discovered connection between bosonization and duality
transformations \abelbosedual, together with the observation
that duality transformations are not intrinsically restricted to $d=1$
dimensions \damgaardrefs. We confine our
attention here to the case of the `abelian' bosonization of a single
Dirac fermion, although we expect that a higher-dimensional
generalization of nonabelian bosonization \nonlinearbos\ can
be obtained along the same lines by using the analogous
connection \nonabelbosedual\ between nonabelian bosonization
and nonabelian duality \doq.

Starting with a Dirac fermion in $d+1$ spacetime dimensions,
the dualization approach automatically guarantees the existence
of a bosonized version of the theory, with an explicit expression
for the bosonic action in terms of a path integral over the fermionic
and some auxiliary degrees of freedom. The dual, bosonic, variable
which appears in this bosonic theory is a rank $d-1$, completely
antisymmetric Kalb-Ramond gauge potential, $\Lambda_{\mu_1 \cdots
\mu_{d-1}}$, which is invariant under the gauge freedom $\Lambda
\to \Lambda + d \omega$, where $\omega_{\mu_1 \cdots \mu_{d-2}}$
is an arbitrary $(d-2)$-form. If the original fermion is coupled
to a $U(1)$ gauge potential, $a_\mu$, through the usual
interaction $\Scl_c = i \psibar \gamma^\mu \psi \; a_\mu$, then
we find that $\Lambda$ necessarily couples through the interaction
term $\epsilon^{\mu_1 \cdots \mu_{d+1}} \partial_{\mu_1} a_{\mu_2}
\Lambda_{\mu_3 \cdots \mu_{d+1}}$.

Although the bosonic theory we are led to in this way
is guaranteed to exist, it is not
required to have many of the usual properties that we tend to take
for granted in the $d=1$ case, such as locality. In order to
investigate the properties of the bosonic theory in more detail
we explicitly perform the functional integrals which define the
bosonic action in the limit that the fermion mass, $m$, is
much larger than the momenta of the external fields. This allows
us to systematically determine the form for the bosonic action
as a series in $1/m$. We find the leading term in this expansion,
and show that although the result is nonlocal when $d \ge 3$,
it turns out to be local for the special case $d= 2$.

\ref\anyons{J.M. Leinaas and J. Myrheim, \nci{37}{77}{1}.}
\ref\anyonCS{D.P. Arovas, R. Schrieffer, F. Wilczek and A. Zee,
\npb{251}{85}{117}.}
The $d=2$ special case is interesting for a number of reasons,
besides the locality of the bosonic action, due to the possibility
in this instance of fractional statistics, or anyons \anyons. These
have a now-standard representation in terms of fermions coupled to
a statistical Chern-Simons gauge field \anyonCS, and by following
this field through the bosonization process we
derive here a bosonized formulation for these particles.

\ref\krduality{H. Nicolai and P.K. Townsend, Phys. Lett. {\bf B98} (1981) 257;
M. Duff, in {\it Superspace and Supergravity} ed. by
S.W. Hawking and M. Rocek, (Cambridge University Press, 1981);
E.S. Fradkin and A.A. Tseytlin, Ann. Phys. {\bf 162} (1985) 31;
J.M.F. Labastida, Phys. Lett. {\bf B171} (1986) 377;
C.P. Burgess and A. Kshirsagar, \npb{324}{89}{157};
J.D. Brown, C.P. Burgess, A. Kshirsagar, B.F. Whiting and J.W. York,
\npb{328}{89}{213}.}
Finally, we address a potential conundrum. Aficionados of duality
will recognize that a rank $d-1$ Kalb-Ramond field in $d$ space
dimensions is equivalent to a purely scalar field with derivative
interactions \krduality. Given these derivative couplings, one might worry
that the scalar degree of freedom should be much lighter than the
fermion mass, $m$, and so not properly reproduce the properties
of the fermionic theory. Similarly, one might worry that the couplings
between $\Lambda_{\mu_1 \cdots \mu_{d-1}}$ and the gauge potential,
$a_\mu$, might cause the
scalar to be `eaten' by the gauge potential, and to thereby {\it
always} give this field a mass. We show how this conundrum gets
resolved by explicitly performing this duality, where we find
that the scalar completely decouples from the other external fields.

\ref\frohlich{J. Fr\"ohlich, R. G\"otschmann and P.A. Marchetti,
Z\"urich-Padova preprint DFPD 94/TH/36 (hep-th/9406154).}
While this work was in progress we received Ref. \frohlich, which
takes a somewhat similar point of view to the one taken here.

\section{The Bosonization Algorithm}

We take as our starting point a free Dirac fermion, $\psi$, in
$D = d+1$ spacetime dimensions. We consider explicitly the
relativistic case, but our methods apply equally well for
nonrelativistic systems. (More details concerning the
treatment of nonrelativistic systems may also be found in
Ref. \frohlich, since these were the principal applications of
this reference.) We take the fermionic lagrangian density to be:
\label\fermionlagrangian
\eq
\Scl_\ssf = - \psibar \, [ \dsl + m + J_i M_i ] \psi,
\eeq
where $M_i$ represents the complete set of Dirac matrices that
is appropriate to the chosen dimension of spacetime.\foot\diraceg{
For instance, when $D = 4$ we have $M_i = 1,\gamma_5, \gamma^\mu,
\gamma^\mu \gamma_5$ and $\gamma^{\mu\nu}$.} The $J_i$ are a collection
of external fields, and it is the response of the system to these
fields which we wish to compute. (For practical applications below,
we take $J_i$ to be an applied electromagnetic field,
$a_\mu$, for which $\psibar M_i \psi = i \psibar \gamma^\mu \psi$.)

To bosonize we follow Ref. \abelbosedual\ and first enlarge
the fermion theory by gauging the global
$U(1)$ symmetry $\psi \to e^{i \theta} \, \psi$, while constraining
the field strength of the corresponding gauge potential, $A_\mu$,
to vanish. The lagrangian of this enlarged theory is:
\label\gaugelagrangian
\eq
\Scl_\ssg = \Scl_\ssf + i \psibar \gamma^\mu \psi \, A_\mu +
\epsilon^{\mu_1 \cdots \mu_{d+1}} \partial_{\mu_1} A_{\mu_2}
\, \Lambda_{\mu_3 \cdots \mu_{d+1}}.
\eeq
The key point is that this extended theory is precisely equivalent
to the original system described by eq. \fermionlagrangian.
This can be seen by first integrating the lagrange-multiplier
field, $\Lambda_{\mu_1 \cdots \mu_{d-1}}$, and then integrating
over $A_\mu$. Integration over the lagrange-multiplier field
imposes a constraint which ensures that $A_\mu$ is gauge-equivalent
to zero. This forces eq. \gaugelagrangian\ to reduce to
eq. \fermionlagrangian, and establishes the equivalence of these
two theories.

The bosonized result is then obtained by starting from eq.
\gaugelagrangian, but instead integrating out the fields
$\psi$ and $A_\mu$. Only $\Lambda_{\mu_1 \cdots \mu_{d-1}}$ is
left to play the role of the bosonized variable. The bosonized
lagrangian is therefore defined by the following functional
integral (with $J_i=a_\mu$):
\label\bosonlagrangiandef
\eq
\exp\Bigl[ i S_\ssb(\Lambda,a) \Bigr] =
\int \Scd\psi \, \Scd A_\mu \; \exp \left[ i S_\ssg(\psi,\Lambda,a,A)
+ {i \over 2 \, \xi} \int d^Dx \; (\partial^\mu A_\mu )^2 \right].
\eeq
The final term is a gauge averaging term which we have chosen
to use to gauge fix the $A_\mu$ integration.

This last equation, eq. \bosonlagrangiandef, is our
starting point which defines the bosonic theory for
spaces of arbitrary dimension. We next turn to the explicit
evaluation of the integrals.

\section{The Large-$m$ Limit}

Although it is not known how to evaluate the functional integrals of
eq. \bosonlagrangiandef\ in the general case, they can be performed
approximately in certain circumstances. In this section we wish to
evaluate them using an expansion in powers of the inverse fermion mass,
$1/m$. We focus here on the leading behaviour in this limit, but
higher orders in $1/m$ can be obtained in a similar way.

The fermionic functional integration is comparatively simple to
evaluate in the large-$m$ limit since
it leads to a local {\it effective lagrangian} that is dominated
by those interactions which have the lowest scaling dimension.
For a perturbative system such as the one considered here, lowest
scaling dimension reduces to lowest naive dimension, leading to
a very simple expression for the result:
\label\fermionintegral
\eq \eqalign{
\exp\Bigl[ i \Gamma_\ssf(A) \Bigr] &\equiv \int \Scd\psi \exp
\left\{ -i \int d^Dx \; \psibar \Bigl[ \gamma^\mu \left(
\partial_\mu - i A_\mu \right) + m \Bigr] \psi \right\} \cr
&= \exp \left\{ {i \over 2} \int d^Dx \; A_\mu \Pi^{\mu \nu}_D
A_\nu + \cdots \right\}, \cr}
\eeq
where the ellipses denote terms whose coefficients involve
additional powers of $1/m$. We drop here, and elsewhere,
any irrelevant field-independent multiplicative constants.
\foot\technical{A technical aside: The fermionic
functional integral diverges, and we choose to regularize these
divergences using dimensional regularization. All such divergences
must be renormalized, as usual, and we do so using $\ss \MSbar$.
This scheme has the advantage of making it simple to track the
$\ss m$ dependence of the results.} Notice that the condition of
lowest dimension implies that, for all spacetime dimensions,
the lowest-dimension operator is just quadratic in the applied
field, $A_\mu$.\foot\quadratic{This observation is also made
for more complicated systems in Ref. \frohlich.} This permits
the explicit calculation of the subsequent integrals over the
potential $A_\mu$.

\ref\emresponse{W. Heisenberg and H. Euler, {\it Z. Physik}
{\bf 98} (1936) 714;
J. Schwinger, \pr{82}{51}{664}; \pr{93}{54}{615};
\pr{94}{54}{1362}; \pr{128}{62}{2425}.}
Calculations of the response of a Dirac fermion to applied
electromagnetic fields have a distinguished history \emresponse,
and the corresponding vacuum polarization,
$\Pi_D^{\mu\nu}$, takes different forms
in the cases $D=3$ (\ie\ $d=2$) and $D \ge 4$ ($d \ge 3$).
This is the origin of the differences in the properties of
their bosonized forms. For $D\ge 4$ one has:
\label\fourdresult
\eq
\Pi_D^{\mu\nu} = k_D \; (\Square \, \eta^{\mu\nu} - \partial^\mu \partial^\nu)
\eeq
where $k_D$ is a $D$-dependent constant. (For even $D$,
$k_D$ is also a divergent
constant, and so must be renormalized. As a result, the only unambiguous
conclusions that can be drawn about its form is how $k_D(\mu)$ runs
as the renormalization point, $\mu$, changes.)

\ref\threedemresponse{A.N. Redlich, \prd{29}{84}{2366}.}
For $D=3$ ($d=2$) there is an operator which has lower scaling
dimension than that of eq. \fourdresult, and its coefficient is
known to be finite \threedemresponse:
\label\threedresult
\eq
\Pi_3^{\mu\nu} = k_3 \; \epsilon^{\mu\lambda\nu} \partial_\lambda,
\eeq
with $k_3 = \hbox{sign}(m)/( 8 \pi^2)$.

As applied to our starting point, eq. \bosonlagrangiandef, for
the case of an applied electromagnetic field ($J_i = a_\mu$)
we have to leading order in $1/m$:
\label\Aintegral
\eq \eqalign{
\exp\Bigl[ i S_\ssb(\Lambda,a) \Bigr] &= \int \Scd A_\mu \;
\exp \Bigl\{ i \Gamma_\ssf(a + A) \cr
& \qquad\qquad\qquad \qquad + i \int d^Dx \; \Bigl[
\epsilon^{\mu_1 \cdots \mu_{d+1}} \partial_{\mu_1} A_{\mu_2}
\, \Lambda_{\mu_3 \cdots \mu_{d+1}} +
{1 \over 2 \, \xi} (\partial^\mu A_\mu )^2 \Bigr] \Bigr\} \cr
&= \exp\Biggl\{ -{i\over 2} \int d^Dx \; \Bigl[ \Omega_\mu \,
(\hat{\Pi}_D^{-1})^{\mu\nu} \, \Omega_\nu + 2 \, \Omega_\mu \,
(\hat{\Pi}_D^{-1})^{\mu\lambda} \, (\Pi_D)_{\lambda\nu} \,
a^\nu \cr
& \qquad\qquad\qquad\qquad
+ a_\mu (\Pi_D)^{\mu\lambda} \, (\hat{\Pi}_D^{-1})_{\lambda\rho}
(\Pi_D)^{\rho \nu} a_\nu - a_\mu (\Pi_D)^{\mu\nu} a_\nu \Bigr]
\Biggr\}.  \cr}
\eeq
Here $(\hat{\Pi}_D)^{\mu\nu} = (\Pi_D)^{\mu\nu} + {1 \over \xi}
\; \partial^\mu \partial^\nu$ represents the gauge-fixed operator,
and $\Omega_\mu$ is the Hodge dual of the field strength for
$\Lambda_{\mu_1 \cdots \mu_{d-1}}$; \ie\ $\Omega^\mu \equiv
\epsilon^{\mu \mu_2 \cdots \mu_{d+1}} \, \partial_{\mu_2}
\Lambda_{\mu_3 \cdots \mu_{d+1}}$.

As written, eq. \Aintegral\ holds even for nonrelativistic
$\Pi_D^{\mu\nu}$. This equation simplifies significantly for
the relativistic vacuum polarizations of eqs. \fourdresult\
and \threedresult, however. Using the following identities
(which hold equally well for both $D=3$ and $D \ge 4$):
\label\identities
\eq
(\hat{\Pi}_D^{-1})^{\mu\lambda} \, (\Pi_D)_{\lambda\nu} =
\delta^\mu_\nu - { \partial^\mu \partial_\nu \over \Square} ;
\qquad
(\Pi_D)^{\mu\lambda} \, (\hat{\Pi}_D^{-1})_{\lambda\rho}
(\Pi_D)^{\rho \nu} = \Pi_D^{\mu\nu} ,
\eeq
eq. \Aintegral\ reduces to
\label\relAintegral
\eq
S_\ssb(\Lambda,a) =  -{1\over 2}
\int d^Dx \; \left[ \Omega_\mu \, (\hat{\Pi}_D^{-1})^{\mu\nu} \,
\Omega_\nu + 2 \, \Omega_\mu \, \left(\eta^{\mu\nu} - { \partial^\mu
\partial^\nu \over \Square} \right) \, a_\nu \right].
\eeq

\subsection{The Case $d \ge 3$}

Consider first the case $D\ge 4$ ($d\ge 3$). Explicitly
inverting $\hat{\Pi}_D^{\mu\nu}$ gives:
\label\fourdinverse
\eq
(\hat{\Pi}_D^{-1})^{\mu\nu} = {1 \over k_D} \; \left[ \eta^{\mu\nu}
+ (\xi \, k_D -1) {\partial^\mu \partial^\nu \over \Square} \right]
{1 \over \Square}.
\eeq
This, together with the identity $\partial_\mu \Omega^\mu = 0$,
leads to the following expression for the bosonic action:
\label\fourDAintegral
\eq
S_\ssb(\Lambda,a) =  -  \int d^Dx \; \left[ {1\over 2 k_D}
\Omega_\mu \, {1 \over \Square} \, \Omega^\mu + \Omega_\mu \,
a^\mu \right].
\eeq
Notice, as advertised, the nonlocality of the $\Omega_\mu \;
\Square^{-1} \Omega^\mu$ term.

Since the integral over $\Lambda_{\mu_1 \cdots \mu_{d-1}}$ is
gaussian, it may be performed explicitly to give the large-$m$
limit of the fermion integration of the original theory. We do
not perform this integral here, as we do so in a later section,
while dualizing the $(d-1)$ form to a scalar field.

\subsection{The Case $d = 2$}

The special case $D=3$ ($d=2$) proceeds along the same lines.
The inverse vacuum polarization is
\label\threedinverse
\eq
(\hat{\Pi}_3^{-1})^{\mu\nu} = {1 \over k_3} \; \left[
\epsilon^{\mu\lambda\nu} \partial_\lambda - \xi \, k_3
{\partial^\mu \partial^\nu \over \Square} \right] {1 \over \Square}.
\eeq
Inserting this expression into eq. \relAintegral\ then gives the
{\it local} Chern-Simons result
\label\threeDAintegral
\eq
S_\ssb(\Lambda,a) =  -  \int d^3x \; \epsilon^{\mu\lambda\nu}
\left[ {1 \over 2 k_3} \Lambda_\mu \partial_\lambda \Lambda_\nu +
a_\mu \partial_\lambda \Lambda_\nu \right].
\eeq

As for the case of $D \ge 4$, it is straightforward to verify the
equivalence of the bosonic theory we have obtained with the original
fermionic one, by performing the gaussian $\Lambda_\mu$ integration.

\subsection{Anyons}

We pause here to briefly consider the dualization of
anyons in $D=3$ ($d=2$) dimensions. A free anyon having a statistical
phase, $\theta$, has a standard representation \anyonCS\ in terms
of a fermion coupled to a dummy statistics gauge field, $s_\mu$,
with an action of the form:
\label\freeCSanyonlagrangian
\eq
\Scl_{\rm anyon} = - \psibar \Bigl[ \gamma^\mu ( \partial_\mu
 -i a_\mu -i s_\mu ) + m \Bigr] \psi - { 1 \over 2 \theta} \;
\epsilon^{\mu\lambda \nu} s_\mu \partial_\lambda s_\nu.
\eeq
We can clearly bosonize this system, in the limit of large $m$,
as in the previous sections. To apply these results one must simply:
($i$) shift the external electromagnetic field, $a_\mu$, by $a_\mu
\to a_\mu + s_\mu$, and ($ii$) add the Chern Simons action for $s_\mu$.
We find in this way a formulation for long-wavelength ($m \to \infty$)
effects of the anyons, in terms of purely bosonic fields:
\label\bosonanyons
\eq
\Scl_{\rm anyon} =  - \epsilon^{\mu\lambda\nu} \left[ {1 \over 2 k_3}
\Lambda_\mu \partial_\lambda \Lambda_\nu + (a_\mu + s_\mu)
\partial_\lambda \Lambda_\nu +  { 1 \over 2 \theta} \;
s_\mu \partial_\lambda s_\nu \right].
\eeq

\section{Dualizing to a Scalar Variable}

We next turn to the dualization of the antisymmetric Kalb-Ramond
fields, $\Lambda_{\mu_1 \cdots \mu_{d-1}}$, into derivatively-coupled
scalars --- a
process which is possible for all $d$ \krduality. This is a useful
exercise for many reasons, not least because we have more
intuition about the behaviour of scalars than we do about Kalb-Ramond
fields. In particular, it provides a simple way of evaluating
the functional integrals over $\Lambda_{\mu_1 \cdots \mu_{d-1}}$
for comparison with the known fermionic result (in the large-$m$
limit).

A puzzle immediately presents itself once it is realized that the
scalar which is dual to a Kalb-Ramond field must automatically be
derivatively coupled, since how can it be possible that such a
boson can reproduce the properties of a very massive fermion?
Furthermore, since $\Lambda_{\mu_1 \cdots \mu_{d-1}}$ typically
couples to an external electromagnetic potential, $a_\mu$, we
might also expect the bosonic theory to always describe a medium
for which electromagnetic gauge invariance is spontaneously broken
(such as in a superconductor). While this may happen for some
systems, it should not appear as a general result of the large-$m$
limit. We resolve this puzzle here by explicitly performing this duality
transformation, thereby constructing the dual scalar theory.

In this case dualization is based on recognizing that the functional
integral over $\Lambda_{\mu_1 \cdots \mu_{d-1}}$ can be rewritten
as an integral over its field strength, $\Omega^\mu = \epsilon^{
\mu \mu_2 \cdots \mu_{d+1}} \partial_{\mu_2} \Lambda_{\mu_3 \cdots
\mu_{d+1}}$, subject to the constraint that $\partial_\mu \Omega^\mu
\equiv 0$. For an arbitrary functional, $F(d\Lambda)$, of the field
strength, $d\Lambda$, we therefore write:
\label\dualization
\eq
\int \Scd\Lambda \; F(d\Lambda) = \int \Scd\Omega \; F(\Omega)
\; \delta(\partial \cdot \Omega) = \int \Scd\Omega \; \Scd\varphi
\; F(\Omega) \exp\Bigl[ i \int d^Dx \, \varphi \partial_\mu
\Omega^\mu \Bigr].
\eeq
The dual theory is then obtained by performing the (now
unconstrained) integration over $\Omega_\mu$. Since for the
cases of interest here, the integrals over $\Omega_\mu$ are
all gaussian, they may be performed explicitly, and they give
the same result for all $D$, including $D=3$:
\label\scalarform
\eq
S_\ssb(\varphi,a) = \hf \int d^Dx \; \Bigl[ a_\mu \,
\Pi_D^{\mu\nu} \, a_\nu + {1 \over \xi} \; \varphi \, \Square
\, \Square \, \varphi \Bigr].
\eeq

Although the $\varphi$ action turns out to be local, it is unorthodox
in that it contains higher derivatives. As a result, $\varphi$ does
not propagate like a normal scalar. This turns out to be largely
irrelevant in this formulation of the theory, however, since $\varphi$
also completely decouples from the low-energy electromagnetic
potential. Its functional integral may therefore be completely
ignored, contributing as it does just a field-independent
overall constant. Notice also that the remaining $a_\mu$-dependent
term is precisely the large-$m$ limit of the fermionic theory
(\cf\ eq. \fermionintegral), allowing us to see in this way
that the decoupling of the scalars, $\varphi$, is just what
is required to reproduce the original fermionic result.

\section{Conclusions}

In summary, we have presented here a way of bosonizing fermions
in arbitrary dimensions. When applied to a system of space
dimension $d$, and so spacetime dimension $D=d+1$, it leads to
a bosonic variable which is a rank $(d-1)$ Kalb-Ramond field.
When specialized to $D=1+1$ dimensions, this construction
produces a scalar field, and reproduces all of the known results
of abelian and nonabelian bosonization \abelbosedual,
\nonabelbosedual.

We give the action for the bosonic form of the theory in general
as an explicit integral over the fermionic, as well as some auxiliary,
degrees of freedom.  These integrals can be performed in the limit
of very massive fermions, for which an expansion in powers of $1/m$
can be set up. The leading term for this expansion gives gaussian
integrals which can be performed explicitly. Although the
ability to perform all of the integrals in this case obviates much of
the necessity for bosonization, the real utility of the method
comes in once interactions are included, since these can be
recast into bosonic form and analyzed there. (A very nice example
along these lines is given in Ref. \frohlich, where a contact
four-fermion interaction becomes a gaussian integration when written
in the bosonic formulation.)

We find that in the general case, the bosonic theory that is obtained
in this way is nonlocal. The case $D=3$ ($d=2$) is special, however,
and leads to a local Chern-Simons lagrangian (in the large-$m$ limit)
for the bosonic field $\Lambda_\mu$. Anyons may also be rewritten
using our techniques, through the artifice of introducing a
Chern Simons statistics field.

Finally, we show how the bosonic theory can again be dualized
and thereby reexpressed (for any $d$) as a scalar field theory.
The resulting scalar has a higher-derivative propagator, and
decouples from the other low-energy degrees of freedom in all
the cases we considered.

\bigskip
\centerline{\bf Acknowledgments}
\bigskip

This research was partially funded by the N.S.E.R.C.\ of Canada,
les Fonds F.C.A.R.\ du Qu\'ebec, the Norwegian Research Council, and
the Swiss National Foundation.

{\baselineskip 15pt
\listrefs
}

\bye